\documentclass[12pt]{article}

\usepackage{a4}
\usepackage{psfig}
\usepackage{amssymb}

\newcommand{\half}{\frac{1}{2}}
\newcommand{\be}{\begin{equation}}
\newcommand{\ee}{\end{equation}}
\newcommand{\bean}{\begin{eqnarray*}}
\newcommand{\eean}{\end{eqnarray*}}
\newcommand{\bea}{\begin{eqnarray}}
\newcommand{\eea}{\end{eqnarray}}
\newcommand{\intk}{\int\frac{d^3k}{(2\pi)^3}}
\newcommand{\bra}{\langle}
\newcommand{\ket}{\rangle}

\newcommand{\dxp}{d^3 x'\,}
\newcommand{\dxpp}{d^3 x''\,}
\newcommand{\vecnul}{{\bf 0}}
\newcommand{\vecp}{{\bf p}}
\newcommand{\veck}{{\bf k}}
\newcommand{\vecx}{{\bf x}}

\newcommand{\gm}{\gamma}
\newcommand{\dl}{\delta}
\newcommand{\ep}{\epsilon}

\newcommand{\lm}{\lambda}

\newcommand{\om}{\omega}
\newcommand{\Om}{\Omega}

\newcommand{\Lm}{\Lambda}

\makeatletter
\def\sech{\mathop{\operator@font sech}\nolimits}
\makeatother

\begin{document}

\title{
\vskip -100pt
{\begin{normalsize} 
\mbox{} \hfill THU-96-37\\
\mbox{} \hfill ITFA-96-43\\
\mbox{} \hfill October 1996\\
\vskip  70pt
\end{normalsize}}
Finiteness of Hot Classical Scalar Field Theory 
and the Plasmon Damping Rate }
\author{
Gert Aarts$^a$\thanks{email: aarts@fys.ruu.nl} 
\addtocounter{footnote}{1} 
and 
Jan Smit$^{a,b}$\thanks{email: jsmit@phys.uva.nl}\\ 
\normalsize
{\em $\mbox{}^a$Institute for Theoretical Physics, Utrecht University}\\
\normalsize
{\em            Princetonplein 5, 3508 TA Utrecht, the Netherlands}\\
\normalsize
{\em $\mbox{}^b$Institute for Theoretical Physics, University of Amsterdam}\\ 
\normalsize
{\em            Valckenierstraat 65, 1018 XE Amsterdam, the Netherlands}
\normalsize
}
%\date{\today}

\maketitle

\renewcommand{\abstractname}{\normalsize Abstract}
\begin{abstract}
  \normalsize 
We investigate the renormalizability of the classical $\phi^4$ theory at 
finite temperature.
We calculate the time-dependent two point function to two loop order and show 
that it can be rendered finite by the counterterms of the
classical static theory. As an application the classical
plasmon damping rate is found to be $\gamma = \lambda^2 T^2/1536 \pi m$.
When we use the high temperature expression for $m$ given by dimensional 
reduction, the rate is found to agree with the quantum mechanical result.

\end{abstract}

\newpage
\section{Introduction.}

In recent years there has been 
an increasing interest in 
real-time phenomena in finite temperature quantum field theory. 
The complexity of the phenomena often calls for numerical simulations.
Because it is extremely difficult to simulate
quantum mechanical real-time quantities, approximations have to 
be made. In particular the classical approximation has been used 
to simulate diffusion of the Chern-Simons number in gauge theories 
[1-4]
and the dynamical properties of the electroweak phase transition 
\cite{MoTu96}.
It has been argued that the classical description can be accurate 
for physical quantities which are mainly determined by low 
momentum modes of the theory, because these modes will be highly 
occupied at high temperatures.
However, a clear demonstration is complicated by the fact that there 
may be modes with momentum higher that the temperature. Excluding 
these by a cutoff in an effective field theory allows for a 
controlled derivation, but it leads to very complicated dynamics 
with nonlocal noise terms \cite{GrMu96}. It is much simpler to use
cutoffs much higher than the temperature and deal with dynamics 
which is local at this cutoff scale. In this letter we shall 
study the limit of infinite cutoff of the (noiseless) classical 
theory in perturbation theory, taking the $\phi^4$ theory as an 
example.

Classical time dependent correlation functions are determined by 
solving classical equations of motion for arbitrary initial 
conditions, and performing a Boltzmann weighted sum over these 
initial conditions. For time independent correlation functions, 
this prescription reduces to calculations in a three dimensional,
euclidean field theory. In fact, this static theory is of the 
same form as the dimensional reduction approximation in finite 
temperature field theory \cite{AmKra95,TaSm96}. In one method of 
dimensional reduction, the values of the coupling constants are 
precisely specified by a matching procedure between the four and 
the three dimensional theory \cite{Kaea96a}. In the classical 
theory the coupling constants are apriori arbitrary. The 
dimensionally reduced theory has been used extensively for the 
study of the electroweak phase transition \cite{Kaea96b}.

The euclidean theory coming from dimensional reduction is 
superrenormalizable and therefore, time independent correlation 
functions can be made finite by merely a mass renormalization. It 
is not apriori clear that time dependent correlation functions 
can be made finite by the same renormalization. In 
\cite{BoMcLeSmil95} it is shown for a $\phi^4$-theory that the 
time dependent two point function can be made one loop finite by 
supplying the mass counterterm of the static theory. However, at 
two loops the situation is different because then the correlation 
functions  of the static theory mix with time dependent Green 
functions. We shall study this question by calculating the 
classical time dependent two point function to two loop order. 

We will solve the equations of motion by applying naive 
perturbation theory, which leads to resonant or secular terms in 
each step of the expansion. These can be avoided by adding 
counterterms to eliminate the secular terms. We will show that 
the counterterms needed to make the theory finite can be chosen 
such that they also partially take care of the resonances.

We also calculate the classical plasmon rate and show that it is 
equal to the quantum mechanical expression at high $T$, when the 
connection is made with dimensional reduction.

\section{Definition of the correlation functions.}
We consider a scalar field theory at temperature $T = 1/\beta$, 
with the following hamiltonian 
\[  
H = \int d^3x \left(\half \pi^2 + \half (\nabla \phi)^2 + \half \mu^2
\phi^2 + \frac{1}{4!}\lambda\phi^4 + \epsilon \right).
\]   
Classical real time $n$ point functions are 
defined in the following way 
\[ \bra \phi(\vecx_1, t_1)\ldots\phi(\vecx_n, t_n)\ket =
\frac{1}{Z}
\int {\cal D}\pi {\cal D}\phi e^{-\beta 
H(\pi,\phi)}\phi(\vecx_1, t_1)\dots\phi(\vecx_n, t_n),\]
where
\[ Z = \int  {\cal D}\pi {\cal D}\phi e^{-\beta H(\pi, \phi)},\] 
and $\phi(\vecx,t)$ is the solution of the equations of motion
\bean
\dot{\phi}(\vecx,t) &=& \pi(\vecx,t),\\
\dot{\pi}(\vecx,t) &=& 
\nabla^2\phi(\vecx,t) -\mu^2\phi(\vecx,t)-\frac{\lambda}{3!}\phi^3(\vecx,t),
\eean
with initial conditions
\[ \phi(\vecx, t_0) = \phi(\vecx),\;\;\;\; \pi(\vecx, t_0) = \pi(\vecx).\]
The integrals over $\pi$ and $\phi$ are performed at $t=t_0$ (summing
over the initial conditions). Since we consider equilibrium correlation
functions, $t_0$ is arbitrary. In particular the two point 
function will only depend on $t_1-t_2$. Choosing infinite volume 
it depends also only on $\vecx_1-\vecx_2$. In the following we 
choose parameters such that the theory is in its symmetric phase.

For equal times $t_1 = \ldots \ = t_n = t_0$
the $\pi$-field decouples and we are left with a three 
dimensional theory for $\phi$. This three dimensional theory is 
superrenormalizable and it can be made finite with merely a mass 
renormalization. The counterterms are determined by the leaf 
diagram and the momentum independent part of the setting sun 
diagram.  Writing
\[
\mu^2 = m^2 -\delta m^2, \;\;\; \delta m^2 =  \lambda m_1^2 
+ \lambda^2 m_2^2,
\] 
we will use 
\bea 
\label{ct1} 
m_1^2 &=&  \half T \intk \frac{1}{\omega_\veck^2} =
\frac{1}{4\pi^2}\, \Lambda T 
- \frac{1}{8\pi}\, mT,\\
m_2^2 &=& -\frac{1}{6} T^2 
\int \frac{d^3k_1}{(2\pi)^3}
\, \frac{d^3k_2}{(2\pi)^3}
\, \frac{d^3k_3}{(2\pi)^3}\,
\frac{(2\pi)^3\delta(\veck_1 + \veck_2 + \veck_3)}
{\omega_{\veck_1}^2\omega_{\veck_2}^2\omega_{\veck_3}^2}\nonumber\\
&=&\mbox{} -\frac{T^2}{32\pi^2}\log(\frac{\Lambda^2}{m^2})
+ \mbox{finite},
\label{ct2}
\eea
with $\Lambda$ the momentum cutoff and $\omega_\veck^2 = 
\veck^2+m^2$. The choice of the finite parts of the counterterms 
implied by the above expressions will turn out to be convenient 
for the moment. With this mass renormalization the equal time 
correlation functions are finite.

The first few terms of a perturbative solution $\phi = \phi_0 + 
\lambda \phi_1 + \lambda^2 \phi_2 + \cdots$ to the equations of 
motion can easily be constructed. The zeroth order solution is 
given by
\[
\phi_0(\vecx,t) = \intk e^{i\veck\cdot \vecx}
\left[\phi(\veck)\cos(\omega_\veck(t-t_0)) +
\frac{\pi(\veck)}{\omega_\veck}\sin(\omega_\veck(t-t_0))\right],
\]
where $\phi(\veck)$ and $\pi(\veck)$ are the Fourier components 
of the initial conditions. Integration over the initial 
conditions gives the free two point function
\[
\bra \phi_0(\vecx,t) \phi_0(\vecx',t')\ket_0 
\equiv S_0(\vecx-\vecx',t-t')
=T\intk e^{i\veck \cdot (\vecx-\vecx')}
\frac{\cos(\omega_\veck(t-t'))}{\omega_\veck^2}.
\]
Imposing the initial conditions $\phi_1 = \dot\phi_1 = \phi_2 
= \dot\phi_2 = \cdots = 0$ the higher order terms can be found with 
the help of the retarded Green function
\[
G^R_0(\vecx-\vecx',t-t') = \intk e^{i\veck \cdot(\vecx-\vecx')}
\theta(t-t')\frac{\sin(\omega_\veck(t-t'))}{\omega_\veck}.
\]
\begin{figure}
\centerline{
\hbox{
\psfig{figure=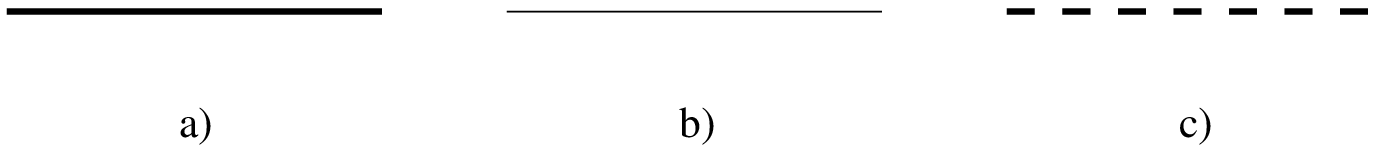,width=4.5in}
}}
\caption{Propagators, a) $S_0(\vecx -\vecx', t-t')$, b) $S_0(\vecx -\vecx', 0)$,
c) $G_0^R(\vecx -\vecx', t-t')$.}
\label{fig1}
\end{figure}
The solution to second order is given by
\bean
\phi(\vecx, t) &=& \phi_0(\vecx,t) \\
&&\mbox{} - \lambda \int dt'\int \dxp G^R_0(\vecx-\vecx',t-t')\left[
\frac{1}{3!}\phi_0^3(\vecx',t') - m_1^2\phi_0(\vecx',t')\right]\\
&&\mbox{} + \lambda^2 \int  dt' \int \dxp G^R_0(\vecx-\vecx', t-t')
m_2^2\phi_0(\vecx',t') \\
&&\mbox{} + \lambda^2 \int dt'\int \dxp \int dt''\int \dxpp
G^R_0(\vecx-\vecx',t-t')G^R_0(\vecx'-\vecx'',t'-t'') \\
&&\;\;\;\;\;\;\;\;\left[\frac{1}{2!}\phi_0^2(\vecx',t') - m_1^2\right]
\left[\frac{1}{3!}\phi_0^3(\vecx'',t'') - m_1^2\phi_0(\vecx'',t'')\right],
\eean
where the integrals over $t', t''$ start at $t_0$.  We expressed 
$\mu^2$ in the equation of motion in terms of $m^2$ and $\delta 
m^2$ and expanded in $\lambda$, using the same coefficients 
$m_1^2$ and $m_2^2$ as found earlier for the equal time case. 
Expanding also the Boltzmann weight, all correlation functions 
can be expressed in terms of the free $n$-point functions of 
$\phi_0(\vecx,t)$, which are just products of the two point 
function $S_0$, by Wick's theorem.
The various propagators are given in figure \ref{fig1}.
It is not clear at this point that the static counterterms 
suffice also for making the time dependent correlation functions 
finite, but we shall see in the following that this is indeed the 
case.

\section{First and second order calculation of the two point function.}
 
Using the framework set up in the previous paragraph, 
we calculate the time dependent two point function to first 
order in $\lambda$. The result is 
\bean
&&\bra \phi(\vecx_1, t_1)\phi(\vecx_2, t_2)\ket
= S_0(\vecx_1 -\vecx_2, t_1- t_2)\\
&&\mbox{} -\half \beta \lambda \int d^3x S_0(\vecx_1 -\vecx, t_1-t_0)
(S_0(\vecnul,0)-2m_1^2) S_0(\vecx -\vecx_2, t_0-t_2)\\
&&\mbox{} -\half \lambda \int dt \int d^3x G_0^R(\vecx_1 -\vecx ,t_1-t)
(S_0(\vecnul,0)-2m_1^2) S_0(\vecx - \vecx_2 ,t-t_2) \\
&&\mbox{} \;\;\;\;\;+ (\vecx_1, t_1) \leftrightarrow (\vecx_2, t_2)
\eean
\begin{figure}
\centerline{
\hbox{
\psfig{figure=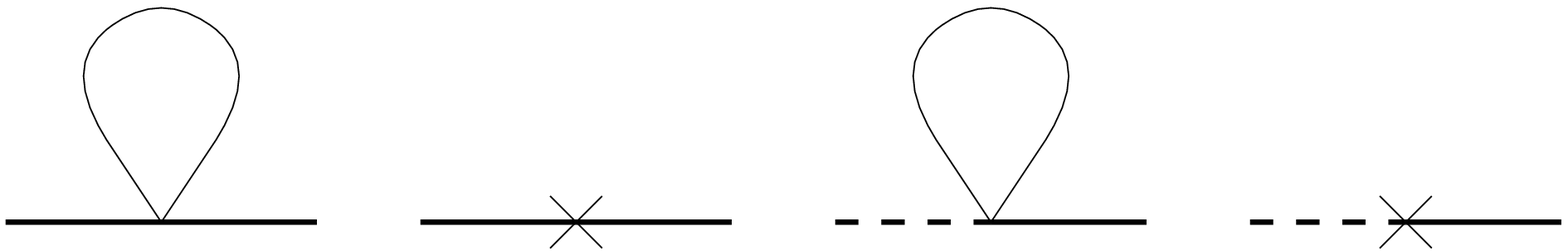,width=5in}
}}
\caption{First order contribution to the two point function, the first order
counterterm is denoted with a cross.}
\label{fig2}
\end{figure}
The corresponding diagrams are given in figure \ref{fig2}. Performing the time 
integral, the correlation function becomes in momentum space 
\bean
S(\vecp,t) &\equiv& \int d^3 x\, e^{-i\vecp \cdot \vecx}\, 
\bra \phi(\vecx, t_1)\phi(\vecnul, t_2)\ket \\
&=& \frac{T\cos(\omega_\vecp t)}{\omega_\vecp^2}\\
&&\mbox{} -\frac{\lambda T }{4}\left(T\int
\frac{d^3k}{(2\pi)^3}\frac{1}{\omega_{\veck}^2} - 2 m_1^2\right)
\frac{1}{\omega_\vecp^4}\left[\omega_\vecp t\sin(\omega_\vecp t) +2\cos(\omega_\vecp t)\right],
\eean
where $t=t_1-t_2$.
We see that the one loop divergence in this time dependent correlation
function is cancelled by the one loop counterterm (\ref{ct1}) of the static 
theory. The particular choice (\ref{ct1}) for the finite terms also cancels
the resonance due to the secular term.
 
We now turn to the second order calculation of the correlation function. 
Using the result for the first order counterterm, which
cancels completely all the one loop contributions, the contribution to second order 
is 
\bean
&&
\bra \phi(\vecx_1, t_1)\phi(\vecx_2, t_2)\ket = 
S_0(\vecx_1 - \vecx_2, t_1- t_2)\\
&\mbox{(a)}& 
\mbox{} + \frac{1}{6}\beta^2 \lambda^2 \int d^3x \, d^3 x'\, 
S_0(\vecx_1 - \vecx, t_1- t_0)S_0^3(\vecx -\vecx',0)
S_0(\vecx' - \vecx_2, t_0-t_2)
\\
&\mbox{(b)}&
\mbox{} +  \beta m_2^2 \lambda^2 \int d^3x\, S_0(\vecx_1 - \vecx, t_1-t_0)
S_0(\vecx - \vecx_2, t_0-t_2)
\\
&\mbox{(c)}&
\mbox{} + \frac{1}{6}\beta \lambda^2 \int dt \int d^3x\, d^3x'\, 
G_0^R(\vecx_1 - \vecx, t_1-t)S_0^3(\vecx - \vecx',t-t_0)\\
&& \;\;\;\;\times S_0(\vecx' - \vecx_2, t_0-t_2) + (\vecx_1, t_1)
\leftrightarrow (\vecx_2,t_2)
\\
&\mbox{(d)}&
\mbox{} + \half \lambda^2 \int dt\, dt'\, \int d^3 x\, d^3 x'\,
G_0^R(\vecx_1 - \vecx, t_1-t)G_0^R(\vecx - \vecx', t-t')\\
&& \;\;\;\;\times
S_0^2(\vecx - \vecx',t-t')S_0(\vecx' - \vecx_2, t'-t_2) + (\vecx_1, t_1)
\leftrightarrow (\vecx_2, t_2)
\\
&\mbox{(e)}&
\mbox{} 
+  m_2^2 \lambda^2 \int dt\int d^3x\, G_0^R(\vecx_1 - \vecx, t_1-t)
S_0(\vecx - \vecx_2, t-t_2) 
\\ 
&&\mbox{} 
\;\;\;\;+(\vecx_1,t_1)\leftrightarrow (\vecx_2,t_2)
\\
&\mbox{(f)}&
\mbox{}
+ \frac{1}{6}\lambda^2 \int dt\, dt'\,\int d^3 x\, d^3 x'\,
G_0^R(\vecx_1  - \vecx, t_1-t)S_0^3(\vecx - \vecx', t-t')\\
&& \;\;\;\;\times G_0^R(\vecx_2 - \vecx',t_2-t').   
\eean
\begin{figure}
\centerline{
\hbox{
\psfig{figure=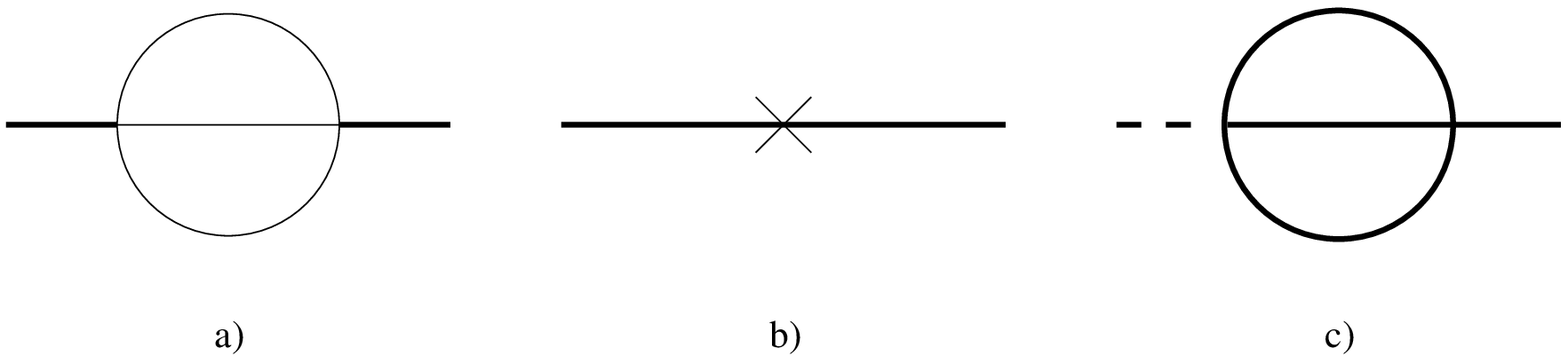,width=4.5in}
}}
\vspace{0.5cm}
\centerline{
\hbox{
\psfig{figure=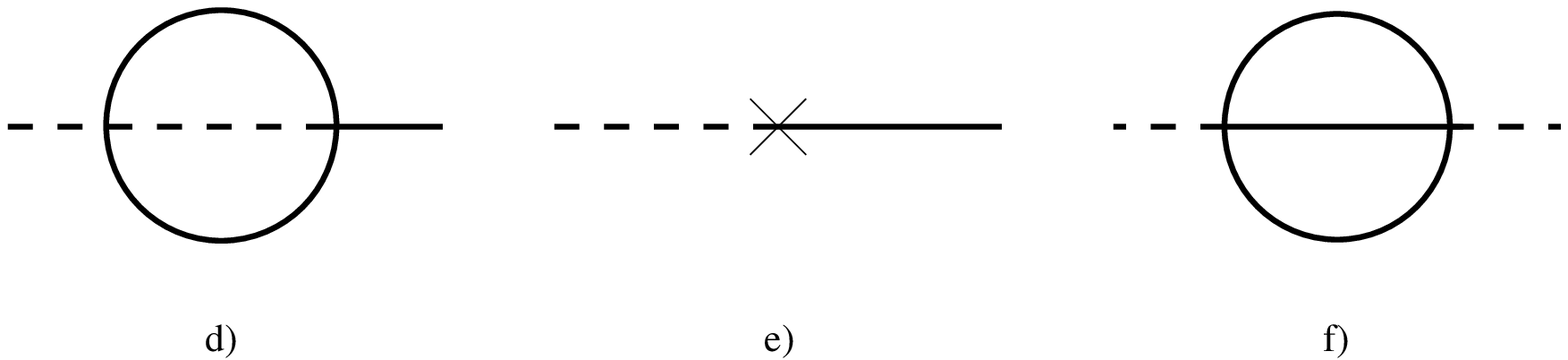,width=4.5in}
}}
\caption{Second order contribution to the two point function; 
here the cross denotes the second order counterterm.}
\label{fig3}
\end{figure}
The diagrams are given in figure \ref{fig3}. Contributions (a) 
and (b) come from the second order expansion of the 
Boltzmannfactor, (c) from a combination of the first order 
expansion of the Boltzmannfactor and the first order solution of 
the equation of motion, (d) and (e) from the second order solution
and (f) from the combination of two first order solutions to the
equations of motion; (a) and (b) are the normal equal time contributions, 
(c), (d), (e) and (f) vanish when $t_1 = t_2 = t_0$. 

\section{Finiteness of the second order contribution.}

We will now see if the counterterm can be chosen in such a way 
that the second order contribution is finite. First we go over to 
three dimensional momentum space. Approximating in expressions 
(a) to (f) the sines and the cosines by 1, we see that there are 
possible linear ((d)) and logarithmic ((a), (c), (f)) divergences 
(actually, the indication of a linear divergence disappears after 
doing the time integrals). Inspection then shows that the 
momentum dependent part is finite, such that only the momentum 
independent part needs to be studied. 
We will take the external momentum 
$\vecp$ to be zero, the internal momenta are labelled as 
$\veck_1, \veck_2$ and $\veck_3 = -\veck_1 - \veck_2$.
Using the fact that the correlation function only depends on 
$t_1-t_2$, we take $t_2 = t_0 = 0$ and $t_1 = t$. This 
immediately sets to zero half of the contribution from (c), (d), 
(e) and the complete contribution from (f). Doing the time 
integrals, the expressions $\omega_{\veck_3} \pm \omega_{\veck_1} 
\pm \omega_{\veck_2} $ (with all four combinations of the 
$\pm$'s) often pop up, and we will denote these by
\[
\Omega = \epsilon_1 \omega_{\veck_1} + \epsilon_2\omega_{\veck_2}
+\epsilon_3\omega_{\veck_3} ,
\]
with $\epsilon_i = \pm 1$. 

Performing the time integrals, using the symmetry of the 
resulting expression to replace 
$3\Omega\epsilon_3\omega_{\veck_3} \to\Omega^2$ and combining 
several contributions, we find the following result, 
\bea
\nonumber
S(\vecnul,t) &=& \frac{T\cos(mt)}{m^2}\\
\nonumber
\mbox{(a+b)}&&
\mbox{} + 
\frac{T\lambda^2\cos(mt)}{m^4}
\left[
\frac{T^2}{6}\int \frac{d^3k_1}{(2\pi)^3}
\frac{d^3k_2}{(2\pi)^3}
\frac{d^3k_3}{(2\pi)^3}\,
\frac{(2\pi)^3\delta(\veck_1 + \veck_2 + \veck_3)}
{\omega_{\veck_1}^2\omega_{\veck_2}^2\omega_{\veck_3}^2}+ m_2^2
\right]\\
\nonumber
\mbox{(d+e)}&&
\mbox{} + 
\frac{T\lambda^2t\sin(mt)}{2m^3}
\left[
\frac{T^2}{6}\int \frac{d^3k_1}{(2\pi)^3}
\frac{d^3k_2}{(2\pi)^3}
\frac{d^3k_3}{(2\pi)^3}\,
\frac{(2\pi)^3\delta(\veck_1 + \veck_2 + \veck_3)}
{\omega_{\veck_1}^2\omega_{\veck_2}^2\omega_{\veck_3}^2}+ m_2^2
\right]\\
\nonumber
\mbox{(c+d)}&&
\mbox{} + 
\frac{T^3\lambda^2}{48}
\int \frac{d^3k_1}{(2\pi)^3}\, \frac{d^3k_2}{(2\pi)^3}\,
\frac{d^3k_3}{(2\pi)^3}\,
\frac{(2\pi)^3\delta (\veck_1+\veck_2+\veck_3)}
{\omega^2_{\veck_1}\omega^2_{\veck_2}\omega^2_{\veck_3}}\\
\label{xxx}
&&\;\;\;\;\times\sum_{\epsilon_1\epsilon_2\epsilon_3}
\frac{1}{\Omega^2 - m^2}
\left[ \frac{t\sin(mt)}{2m} + 
\frac{\cos(\Omega t) - \cos(mt)}{\Omega^2 -m^2} \right].
\eea
In parenthesis we have indicated which diagrams contributed.
Both (a+b) and (d+e) give the same result for the second order 
counterterm, it is given by (\ref{ct2}). 
This is an important result because it shows that the counterterms
of the static theory are sufficient to make the time dependent two point
function two loop finite. With the choice (\ref{ct2}) (again 
in particular the finite part), the above (a+b) and (d+e) are 
zero. The contribution indicated by (c+d) is ultraviolet finite.  
The possible divergences for $\Omega \to \pm m$ are actually not 
present. For example, for $\Omega \to m$,
\[
\frac{1}{\Omega^2 - m^2}\,
\left[ \frac{t\sin(mt)}{2m} + \frac{\cos(\Omega t) - \cos(mt)}{\Omega^2 -
m^2} \right] = \frac{t\sin mt}{8m^3} - \frac{t^2\cos mt}{8m^2} 
+ O(\Omega - m).
\]
Let us record here also the fact that the two loop correction is 
$O(t^4)$ for $t\to 0$.

\section{Damping rate}

As an application we derive an expression for the damping rate 
$\gm$, assuming that the dominant large time behavior of the 
correlation function at zero momentum is given by $Z(T/m^2) 
\exp(-\gm t) \cos((m+\dl)t)$. Since $\gm$ and $Z-1$ are 
$O(\lm^2)$ and $\dl$ at least $O(\lm)$ we can identify $\gm$ from 
the coefficient of $-t\cos (mt)$ in the large time behavior of 
(\ref{xxx}). In terms of
\bean
w(\Om) &=& \frac{\lm^2}{6} \sum_{\ep_1\ep_2\ep_3} \int 
\prod_{j=1}^3 
\left[\frac{d^3 k_j}{(2\pi)^3 2\om_{\veck_j}}\, \frac{T}{\om_{\veck_j}}\right]\\
&&
(2\pi)^4 \dl (\veck_1 + \veck_2 + \veck_3)\,
\dl(\ep_1\om_{\veck_1} + \ep_2 \om_{\veck_2} + \ep_3\om_{\veck_3} - \Om)\,
\eean
we can rewrite the $O(\lm^2)$ contribution to (\ref{xxx}) in the form
\[
S^{(2)}(\vecnul,t) = \int \frac{d\Om}{2\pi}\, w(\Om)\, \frac{1}{\Om^2-m^2}
\left[
\frac{t\sin mt}{2m} + \frac{\cos\Om t - \cos mt}{\Om^2 - m^2}
\right].
\]
The function $w(\Om)$ is even in $\Om\to -\Om$, continuous and 
vanishes like $\Om^{-1}$ at large $\Om$. A large time behavior 
$\propto t\cos mt$ can only come from the singularities at 
$\Om=\pm m$, any other portion of the $\Om$ integral gives 
subdominant behavior. Hence we may conveniently replace $w(\Om) 
\to w(m)$ and evaluate the resulting integral, 
\[
S^{(2)}(\vecnul,t) \to
w(m)\left[\frac{\sin mt}{4m^3} - \frac{t\cos mt}{4m^2} \right],
\]
from which we identify
\[
\gm = \frac{w(m)}{4T}.
\]
The $\ep$'s giving nonvanishing contributions to 
$w(\Om)$ are $\ep_1=\ep_2=1$, $\ep_3=-1$ (cycl) for $\Om > 0$ and 
similar for $\Om <0$. Hence $w(\Om)$ has the form of a phase space 
integral for the scattering $(\vecnul,\Om) + (\veck_3, 
\om_{\veck_3}) \to (\veck_1,\om_{\veck_1}) + (\veck_2,\om_{\veck_2})$ etc. The particles
are distributed according to $T/\om$, which is the high temperature form 
of the quantal distribution $[\exp(\om/T) - 1]^{-1}$. 
Evaluation of $w(m)$ gives for the damping rate
\be
\gm = \frac{\lambda^2 T^2}{1536\pi m}.
\label{gmc}
\ee

\section{Matching to the quantum theory}

The classical theory can be seen as a high temperature 
approximation to the quantum theory. Then the parameters in the 
classical theory are effective parameters which are to be chosen 
such that the classical correlation functions match those in the 
quantum theory as well as possible. For the static theory the 
approximation is equivalent to a form of dimensional reduction 
and this makes it possible to determine the parameters in the 
potential energy part of the hamiltonian in terms of the 
parameters in the quantum theory \cite{TaSm96}. General matching 
rules are obtained in ref.\ \cite{Kaea96a}. For example, matching 
the cosmological constant, mass and coupling parameters to 
one loop order we find 
\bean
\epsilon &=& \bar\epsilon - \frac{\pi^2 T^4}{90} + \frac{\bar m^2 T^2}{24}
- \frac{\bar m^4}{64\pi^2}
\ln\frac{\bar\nu^2}{\nu_T^2}
- \frac{\Lambda^3 T}{6\pi^2}\left(\ln\frac{\Lm}{T}-\frac{1}{3}\right) 
- \frac{\Lm Tm^2}{4\pi^2}
,\\
\mu^2 &=&
 \bar m^2 + \bar\lambda \left(
\frac{T^2}{24} 
- \frac{\bar m^2}{32\pi^2}
%\ln\left(\frac{\bar\nu}{4\pi e^{-\gamma_E}T}\right)^2
\ln\frac{\bar\nu^2}{\nu_T^2}
- \frac{\Lambda T}{4\pi^2} 
\right),\\
\lambda &=& \bar\lambda \left(1 - \frac{3 \bar\lambda}{32\pi^2}\,
%\ln\left(\frac{\bar\nu}{4\pi e^{-\gamma_E}T}\right)^2
\ln\frac{\bar\nu^2}{\nu_T^2}
\right),
\eean 
where $\nu_T \equiv 4\pi\exp(-\gm_E) T$ and less dominant terms as $T\to\infty$
are suppressed.
The quantities on the right hand side are dimensionally 
renormalized parameters in the MS-bar scheme and $\bar\nu$ is the 
corresponding mass scale. Notice how the Rayleigh-Jeans 
divergence is canceled by the $\Lm^3$ counterterm, leaving the 
familiar quantum expression for the pressure
as just a finite renormalization in the classical theory.

The above relation for $\mu^2$ now implies with (\ref{ct1}), 
choosing the MS-bar scale $\bar\nu = \nu_T$,
\[
m^2 = \bar m^2 + \lm \left(
\frac{T^2}{24} - \frac{\bar mT}{8\pi} 
\right).
\]
Inserting this relation in the result (\ref{gmc}) for the damping 
rate gives for large $T$ 
\[
%\gm = \frac{\lambda}{64 \pi}\,\sqrt{\frac{\lambda T^2}{24}},
\gm = \frac{\lambda^{3/2}T}{128\sqrt{6}\, \pi},
\]
which is indeed the plasmon rate obtained in the 
quantum theory \cite{Pa92}.

\section{Discussion}

Among the time-independent correlation functions only the 
two-point function is superficially divergent, because of the 
superrenormalizability. Inspection of the time-dependent 
correlation functions leads to the conclusion that, 
after performing the intermediary time integrals and replacing the  
remaining sines and cosines by 1, the overall degree of divergence
of the $n > 2$-point functions is less than zero. 
Only the $n=2$ case is then 
non-trivial and we have shown to second order in $\lambda$ that 
the time-dependent classical two-point function is finite, after 
taking into account the counterterms of the static theory. We 
conclude therefore (but have of course not given a proof)
that with these counterterms all finite 
temperature time dependent correlation functions in $\phi^4$ 
theory are cutoff independent.

The coupling $\lambda$ was taken to be cutoff independent, 
consistent with the superrenormalizability of the static theory.  
We would like to remark that this static three dimensional theory 
is quite non-generic from the Statistical Physics point of view. 
The coupling $\lambda_3$ of a three dimensional theory has the 
dimension of [mass] and the dimensionless coupling at scale $\nu$ 
is $\lambda(\nu) = \lambda_3(\nu)/\nu$. We may call $\lambda_{B} 
= \lambda(\Lambda)$ the bare coupling and $\lambda_R = 
\lambda(m)$ the renormalized coupling.  Since the system gets 
critical as the ratio $\Lambda/m$ approaches infinity, one might 
expect that the renormalized coupling approaches the nontrivial 
infrared stable fixed point $\lambda^*$ of Ising-like systems 
(the Wilson-Fischer point). This is not the case here, because 
$\lambda_B = \lambda T/\Lambda$ is non-generic: it is is 
automatically tuned to the gaussian fixed point as $\Lambda/m \to 
\infty$. The renormalized coupling $\lambda_R \approx \lambda 
T/m$, which can be arbitrary in the interval $(0,\lambda^*)$.

{}For a clearer understanding of the classical theory as an 
approximation to the quantum theory one would like to go beyond 
the usual dimensional reduction involving only time independent 
quantities. Matching time dependent quantities may introduce at 
two loop order also a finite renormalization of the coefficient 
of $\pi^2$ in the hamiltonian. We have not yet carried out such 
time dependent matching. For a different approach to time 
dependent quantities, based on an expansion in Planck's constant 
$\hbar$, see ref.\ \cite{Bod96}.

The fact that the analytic expression for the classical plasmon 
damping rate combined with parameter matching according to 
dimensional reduction leads to the correct quantum mechanical 
result is quite encouraging for the physically more interesting 
case of the SU(2)-Higgs model (cf.\ ref.\ \cite{ArSoYa96} where 
eq. 4.3 gives the impression that the classical gluon damping
rate is uv-divergent; quantummechanically the gluon damping rate is of
course uv-finite but subtle in the infrared, see e.g. \cite{LeBFl}).
Numerical simulations in the classical SU(2)-Higgs model clearly show the 
damping phenomenon and appear to be compatible with lattice spacing 
independence of the rates \cite{TaSm96b}.\\

Acknowledgement\\
This work is supported by FOM, and by the European Commission TMR 
program ERBFMRX-CT96-0045.

\end{document}